\documentclass[twocolumn,superscriptaddress]{revtex4}
\usepackage[dvipdfm]{graphicx}
\usepackage{amsmath,amssymb,bm}
\allowdisplaybreaks[3]
\begin{document}

\title{Hyperk\"ahler Metrics from Monopole Walls}

\author{Masashi Hamanaka}
\affiliation{Department of Mathematics, Nagoya University, Furo-cho, Chikusa-ku, Nagoya 464-8601 Japan}

\author{Hiroaki Kanno}
\affiliation{Department of Mathematics, Nagoya University, Furo-cho, Chikusa-ku, Nagoya 464-8601 Japan}

\author{Daichi Muranaka}
\affiliation{Department of Mathematics, Nagoya University, Furo-cho, Chikusa-ku, Nagoya 464-8601 Japan}

\begin{abstract}
We present ALH hyperk\"ahler metrics induced from well-separated $\mathrm{SU}(2)$ 
monopole walls which are equivalent to monopoles on $T^2 \times \mathbb{R}$. 
The metrics are explicitly obtained due to Manton's observation
by using 
monopole solutions.
These are doubly-periodic and have the modular invariance 
with respect to the complex structure of the complex torus $T^2$.
We also derive metrics from monopole walls
with Dirac-type singularities.\end{abstract}

\maketitle

\section{Introduction}

Hyperk\"ahler manifolds have played important roles in 
the study of supersymmetric quantum field theories and string theories,
especially, in the context of the string compactifications, duality tests and so on. 
The explicit metric on a compact  hyperk\"ahler manifold is not known except 
trivial examples. On the other hand,
explicit forms of the non-compact hyperk\"ahler metric
have been derived in several ways. Among them the most systematic one is 
the hyperk\"ahler quotient construction 
\cite{HKLR} (see also \cite{GGR}).
In 4-dimensions, the hyperk\"ahler metrics 
satisfy the self-dual Einstein equations and 
arise as the gravitational instanton solutions (see e.g. \cite{EGH}).
These can be classified into some categories:
the ALE, ALF, ALG and ALH spaces \cite{Cherkis}
according to their asymptotic volume growth.

In the context of 3-dimensional gauge theories, hyperk\"ahler metrics
are obtained by considering well-separated monopoles, 
which is due to Manton's observation \cite{Manton} that
the 
dynamics of $k$ well-separated BPS monopoles
can be approximated as a geodesic motion on the asymptotic moduli 
space of the BPS $k$-monopole
if the initial velocities of each monopole are substantially small. 
In this paper we only consider the case with the gauge group  $\mathrm{SU}(2)$ 
or  $\mathrm{U}(2)$.


For a non-periodic BPS $k$-monopole the moduli space can be written as
$\mathcal M_k = \mathbb R^3 \times 
(S^1 \times \widetilde{\mathcal M}_k^0) / \mathbb Z_k$,
where the simply-connected part is denoted by
$\widetilde{\mathcal M}_k^0$,
and the degrees of $\mathbb R^3$ and $S^1$
correspond to the center of mass 
and the gauge degree of global $\mathrm{U}(1)$, respectively.
The dimensions of the $k$-monopole moduli $\mathcal M_k$ is equal to $4k$.
The moduli space  $\mathcal M_k$ can be identified with the moduli space 
of a vacuum on the Coulomb branch 
of the three dimensional $\mathrm{SU}(k)$ super Yang-Mills 
theory with eight supercharges \cite{SeWi}.
The relative moduli space of the 2-monopole $\widetilde{\mathcal M}_2^0$
is known as the Atiyah-Hitchin manifold \cite{AtHi}
which is 
the ALF space with $S^1$ fibration over $\mathbb R^3$.
In the case of well-separated BPS monopoles, 
each monopole carries three moduli of the position 
and a degree of the $\mathrm{U}(1)$ phase modulus.
The latter degree corresponds to the electric charge
and hence we should include the electrical degree of the dyon.
The effective dynamics of the $k$-dyon system can be described by
a sigma model Lagrangian whose target space is the monopole moduli space.
Hence the asymptotic metric of the moduli space of the BPS $k$-monopoles 
can be obtained by calculating the Lagrangian of interactions of
$k$ well-separated BPS monopoles (dyons). The metric is  known as 
the Gibbons-Manton metric \cite{GiMa}.

For a periodic BPS $k$-monopole on $\mathbb R^2 \times S^1$,
which is called the monopole chain \cite{Ward2005, ChKa2001, ChKa2003},
the moduli space is identified with the moduli space 
of a vacuum on the Coulomb branch  
of the four dimensional $\mathrm{SU}(k)$ super Yang-Mills 
theory compactified on  $S^1$ with eight supercharges.
The relative moduli space of the 2-monopole $\widetilde{\mathcal M}_2^0$ 
is the ALG space \cite{ChKa2003}.
Since the periodicity is achieved by a chain of monopoles,
the total energy would diverge due to the infinite number of monopoles. 
However, the Nahm transform can be make well-defined
and the asymptotic metric of the moduli space of monopole chains 
is obtained in the same manner as the non-periodic case \cite{ChKa2002}.
The geodesic motion is also discussed \cite{HaWa, Maldonado, MaWa, Maldonado2}.


\begin{widetext}
\begin{center}
\begin{table}[htbp]
\begin{tabular}{|c|c|c|}\hline
Periodicity of monopole & Super Yang-Mills theory & Asymptotic behavior (4d topology)\\ \hline \hline
$\mathbb R^3$ (non-periodic) & $ \mathcal N =4$ SYM on $\mathbb R^3$ & ALF : $S^1$ fibration on $\mathbb R^3$  \\ \hline
$ S^1 \times \mathbb R^2$ (periodic) & $ \mathcal N =2$ SYM on $\mathbb R^3 \times S^1$ &  ALG :  $T^2$ fibration on $\mathbb C$  \\ \hline
$ T^2 \times \mathbb R$ (doubly-periodic) & $ \mathcal N =1$ SYM on $\mathbb R^3 \times T^2$ & ALH : $T^3$ fibration on $\mathbb R$  \\ \hline
\end{tabular}
\caption{The correspondence of the periodicity of monopole, super Yang-Mills theory and the asymptotic
behavior of hyperk\"ahler metric.}
\end{table}
\end{center}
\end{widetext}

For a doubly-periodic BPS $k$-monopole on $T^2 \times \mathbb R$,
which is called the monopole sheet or wall \cite{Ward2005, Ward2007} (see also \cite{Lee}),
the moduli space is identified with the moduli space 
of a vacuum on the Coulomb branch  
of the five dimensional $\mathrm{SU}(k)$ super Yang-Mills 
theory compactified on  $T^2$ with eight supercharges \cite{HaVa}.
One of the examples of the correspondence between the monopole moduli and
the vacuum moduli of the five dimensional super Yang-Mills 
theory is that the number of the Dirac-type singularity corresponds to that of the matter flavor.
Asymptotically the relative moduli space of the monopole walls is
expected to be the ALH space with $T^{3k-3}$ fibration over $\mathbb R^{k-1}$.
As far as we know, there are no examples of ALH hyperk\"ahler metrics 
in the literature except for the classical metric 
derived from the effective action of
the $ \mathcal N =1$ super Yang-Mills theory 
on $\mathbb R^3 \times T^2$ by Haghighat and Vandoren \cite{HaVa}.
Furthermore, the doubly-periodic monopoles have  
rich properties on the D-brane interpretation, string duality,
and M-theoretic interpretation via the various 
S,T-duality transformations \cite{ChWa}.
Therefore the analysis of the moduli metric would be 
applied to various situation of the corresponding
super Yang-Mills theory, string theory and M-theory.

In this Letter, we derive some asymptotic metrics of the monopole walls on $T^2 \times \mathbb R$
by calculating the effective sigma model Lagrangian of $k$
well-separated BPS walls following Manton's observation. 
In our calculation, the BPS wall is assumed to be 
a doubly-periodic superposition of BPS monopoles in flat three-space.
In the non-periodic direction, the walls are assumed to be well-separated 
to each other compared with the thickness of the monopole wall
so that the fields can be well-approximated by superpositions 
of linearized monopole walls.
The metric computed in this paper is for the case of two identical
nonabelian monopole walls, including 
in the presence of Dirac singularities as well. 
We prove that the induced metrics actually have the modular invariance with
respect to a complex structure $\tau$ of the complex torus $T^2$ in addition to the expected periodicity.
We also present the metrics of monopole walls with Dirac-type
singularities. We see that when we consider $k$ monopole walls the maximum number of singularities
is $2k$ by a simple analysis using the Newton polygon. This
is consistent with the fact that in the $\mathrm{SU}(k)$ super Yang-Mills theory
the number of the matter flavor has the upper bound $2k$. 
This bound is due to the requirement that 
the super Yang-Mills theory is either conformal or asymptotically free.
When the bound is saturated the theory has conformal invariance.

The present metrics would be the most explicit ones of the ALH type
derived from the 
solutions of monopole walls 
including the case with the Dirac-type singularities.  
The symmetry and other properties are consistent with 
the one in the corresponding super Yang-Mills theory \cite{HaVa}.

\section{Setup}

Let $x^\alpha := (x, y, z)$ ($\alpha = 1, 2, 3$) be the coordinates of the 
three dimensional space $T^2 \times \mathbb R$ 
in which $x$ and $y$ are periodic: $x \sim x+1, y \sim y+1$.
The Higgs field $\phi$ and the gauge field $A$ satisfy the Bogomolny equation
\begin{align}
    *D_A\mathit\phi = - F,
\end{align}
where $D_A\mathit\phi := \mathrm d\mathit\phi + [A, \mathit\phi]$
and $F := \mathrm dA + A \wedge A$.
We put a condition that the asymptotic behavior of the Higgs field 
of an $\mathrm{SU}(2)$ solution must be \cite{ChWa}
\begin{align}
    \mathrm{EigVal}\,\phi = 2\pi\mathrm iQ_\pm z 
+ O(1) \quad \text{as}~z \to \pm\infty, \nonumber
\end{align}
where the constants $Q_\pm \in \mathbb Z$ are called the monopole-wall charges.
These are topological charges which are related to the Chern number as
\begin{align}
    Q_\pm = \int_{T_z}c_1(E_\pm) = \frac{\mathrm i}{2\pi}\int_{T_z}\mathrm{tr}\,F_\pm, \nonumber
\end{align}
where $T_z$ is the complex torus at $z$ and $E_\pm$ are 
the line bundles defined at $z\rightarrow \pm \infty$, 
respectively, where the monopole vector bundle $E$ splits 
into eigenvalues of the Higgs field 
as $E\vert_z = E_{+}\oplus E_{-}$ \cite{ChWa}. 
Numerical solutions of the $\mathrm{SU}(2)$ 
monopole walls are studied for $(Q_-,Q_+) = (1, 1)$ 
and $(0, 1)$  \cite{Ward2005, Ward2007}.
The detailed analysis of the boundary conditions and the moduli
space are summarized in \cite{ChWa}.

Let us introduce a standard complex structure 
$\tau := \tau_1 + \mathrm i\tau_2$ ($\tau_1, \tau_2 \in \mathbb{R}$) at the torus $T^2$ 
and introduce a holomorphic coordinate $\xi := x + \tau y$.
The periodicity is now represented by $\xi \sim \xi + m + \tau n$ ($m, n \in \mathbb Z$).
By using the vector notation $\bm x := (\xi, z)$,
the metric on $T^2 \times \mathbb R$ is represented as follows:
\begin{align}
    \mathrm d\bm x \cdot \mathrm d\bm x &:=
    \frac{\nu}{\tau_2}(\mathrm dx^2 + 2\tau_1\mathrm dx\mathrm dy + |\tau|^2\mathrm dy^2) + \mathrm dz^2 \nonumber \\ &\phantom{:}=
    \frac{\nu}{\tau_2}|\mathrm d\xi|^2 + \mathrm dz^2 =: g_{\alpha\beta}\mathrm dx^\alpha\mathrm dx^\beta,
\end{align}
where the volume of the torus is denoted by 
$\nu := \sqrt{\det g}$ ($g := (g_{\alpha\beta})$).
Note that two dimensional metric has three independent components and 
we have traded them with $\tau_1, \tau_2$ and $\nu$.
One of the crucial features of our construction of ALH hyperk\"ahler metrics in the following  is
the invariance of the metric under the modular transformation,
\begin{align}
    \xi    \mapsto \frac{ \xi     }{ c\tau + d   }, \quad
    \tau   \mapsto \frac{a\tau + b}{ c\tau + d   }, \quad
    \tau_2 \mapsto \frac{ \tau_2  }{|c\tau + d|^2}, \label{eqn:Modular transformation}
\end{align}
where $
\left(\begin{smallmatrix} a & b \\ c & d \end{smallmatrix}\right) \in \mathrm{SL}(2, \mathbb Z)$.

\section{Asymptotic Behavior of SU(2) Monopole Walls}

For the purpose of calculating the effective Lagrangian for well-separated monopole walls, 
we should derive the asymptotic form of the $\mathrm{SU}(2)$ monopole walls.
Let us consider $k$ well-separated monopole walls sitting at the points $\bm a_j := (\xi_j, z_j)$ ($j = 1, \cdots, k$).
Here each monopole wall has the charge $(Q_-, Q_+) = (0, 1)$.
It can be regarded as a smooth $\mathrm{SU}(2)$ monopole 
arranged per unit cell \cite{Ward2007}.
(It is not clear that the multi-monopole walls have the moduli of the separations,
however, at least the case of $(Q_-, Q_+) = (1, 1)$ has four-moduli \cite{Ward2005}.)
If the separations $|z_j - z_i|$ are large enough compared with the thicknesses of each monopole wall,
the fields are well-approximated by superpositions of linearized monopole
walls:
\begin{align}
     \phi(\bm x) &= v + \sum_{j = 1}^k \phi^j(\bm x - \bm a_j), \label{eqn:The ansatz of the Higgs field} \\
    A_\xi(\bm x) &= b + \sum_{j = 1}^kA_\xi^j(\bm x - \bm a_j), \quad
    A_z  (\bm x)  = 0                                         ,
\end{align}
where $v$ and $b$ are the vacuum expectation value of the Higgs field
and the background gauge field respectively.
Then we can estimate the asymptotic Higgs field of each monopole wall as a superposition of linearized 't Hooft-Polyakov monopoles
arranged in a finite $(2M + 1) \times (2N + 1)$ rhombic lattice,
\begin{align}
    \phi^j(\bm x) = \frac{1}{4\pi}\sum_{m = - M}^M\sum_{n = - N}^N\frac{- g}{\sqrt{|\xi - m - n\tau|^2 + z^2}},
\end{align}
where $g$ is the magnetic charge of the 't Hooft-Polyakov monopoles.
The summation would diverge in the limit of $M$ and $N$ to infinity.
Such divergence can be avoided in a similar way to the case of
periodic monopoles \cite{ChKa2002}. Namely, 
the asymptotic form of $\phi^j(\bm x)$ for large $|z|$ can be written as
\cite{Linton}
\begin{align}
    \phi^j(\bm x) = \frac{g}{2}|z| - gC_{M, N}, \label{eqn:asymptotic form of the Higgs field}
\end{align}
where $C_{M, N}$ is a positive constant diverging linearly in the limit $M, N \to \infty$.
By substituting (\ref{eqn:asymptotic form of the Higgs field}) into (\ref{eqn:The ansatz of the Higgs field}), we obtain
\begin{align}
\label{8}
    \phi(\bm x) = v_\mathrm{ren} + \frac{g}{2}\sum_{j = 1}^k|z - z_j|,
\end{align}
where $v_\mathrm{ren} := v - kgC_{M, N}$, which can be kept finite with $v$ diverging at the same order as $C_{M, N}$.
We note that the configuration is not localized in the periodic
directions. This implies that the superposition of doubly-periodic monopoles  
is represented as a constituent monopole wall in the asymptotic region.

The asymptotic gauge field can also be derived from the Bogomolny equation 
with (\ref{eqn:asymptotic form of the Higgs field}),
\begin{align}
    A_\xi^j(\bm x) = \frac{\mathrm i\nu g}{8\tau_2}\,\mathrm{sign}(z)\,\bar\xi, \quad A_z^j(\bm x) = 0,
\end{align}
where
\begin{align}
    \mathrm{sign}(z) := \left\{\begin{array}{rl} + 1 & (z > 0) \\ - 1 & (z < 0) \end{array}\right.. \nonumber
\end{align}
In order to make the gauge field doubly-periodic for $\xi \to  \xi + m + \tau n$,
we have to perform appropriate gauge transformations. This means our $\mathrm U (1)$ bundle over
the complex torus is non-trivial. 
Accordingly 
we have to impose the following twisted boundary condition
where the phase  $\theta$ of any functions 
in the fundamental representation of the gauge group shifts as follows;
(cf. Eq. (12) in \cite{Ward2005}):
\begin{align}
    \theta &\mapsto \theta + \frac{\nu g}{4}\,\mathrm{sign}(z)\,y \quad \text{when}\quad \xi \mapsto \xi +    1, \label{eqn:periodic boundary condition 1} \\
    \theta &\mapsto \theta - \frac{\nu g}{4}\,\mathrm{sign}(z)\,x \quad \text{when}\quad \xi \mapsto \xi + \tau. \label{eqn:periodic boundary condition 2}
\end{align}

For later convenience we introduce the following pair of the harmonic function 
and the Dirac potential on $T^2 \times \mathbb R$ 
\begin{align}
    u(z) = \frac{1}{2}|z| - C_{M, N}, \quad w(\bm x) = \frac{\mathrm i\nu}{8\tau_2}\,\mathrm{sign}(z)\,\bar\xi,
    \label{eqn:The harmonic function and the Dirac potential in doubly-periodic space}
\end{align}
which satisfy $u(z) = u(- z)$ and $w(\bm x) = w(- \bm x)$. 
Note that 
$u(z)$ is a harmonic function on $\mathbb R$ with $\delta$-function source at the origin.

\section{Asymptotic Metric from SU(2) Monopole Walls}

As mentioned in the introduction, the interaction of non-static monopoles
involves not only the relative coordinates
but also the relative phases.
The relative phase factor gives rise to non vanishing electric charges and hence converts monopoles into dyons.
The interaction term of the Lagrangian can be obtained from the analysis of the forces between BPS monopoles.
The fact that there is no force between well-separated BPS monopoles with the same charge
implies the existence of a long-range interaction caused by the 
Higgs field which becomes massless in the BPS limit. This is also the case for dyons.
Thus the Lagrangian of the $\ell^\mathrm{th}$ monopole wall can be written as
\begin{align}
    L_\ell &= - (g^2 + q_\ell^2)^{1 / 2}\phi(1 - \bm V_\ell^2)^{1 / 2} \nonumber \\ &
           + q_\ell\bm V_\ell \cdot        \bm A  - q_\ell       A_0
           + g  \bm V_\ell \cdot \tilde{\bm A} - g  \tilde A_0,
\end{align}
where $(g^2 + q_\ell^2)^{1 / 2}$, $q_\ell$ and $\bm V_\ell := (\dot\xi_\ell, \dot z_\ell)$ are
the scalar charge, the electric charge and the velocity of $\ell^\mathrm{th}$ monopole wall respectively.
Note that all the particles have the same magnetic charge $g$, 
while the electric charges $q_\ell$ 
may change particle by particle in general.
The first term of the Lagrangian gives rise to the scalar interaction due to the Higgs field. 
The second and the third terms are the ordinary Lorentz force.
The remaining terms describe the dual magnetic interaction to
the electric Lorentz force. The relevant field is the dual potential $(\tilde{\bm A}, \tilde A_0)$ which satisfies $\tilde F = *F$.
The background fields $\phi$, $\bm A$, $A_0$, $\tilde{\bm A}$ and $\tilde A_0$ are generated by the remaining $(k - 1)$ moving dyons,
which can be obtained from the solutions derived in the previous section.
For $j \neq \ell$, the asymptotic fields of $j^\mathrm{th}$ dyonic monopole wall
at rest can be derived in the same way as the non-periodic monopoles,
\begin{align}
    \phi ^j(\bm x) &= (g^2 + q_j^2)^{1 / 2}u(z),
\end{align}
\begin{alignat}{99}
    A_\xi ^j(\bm x) &=   g  w(\bm x), &\quad \tilde A_\xi^j(\bm x) &= - q_jw(\bm x), \nonumber \\
    A_z   ^j(\bm x) &=             0, &\quad \tilde A_z  ^j(\bm x) &=             0, \nonumber \\
    A_0   ^j(\bm x) &= - q_ju(    z), &\quad \tilde A_0  ^j(\bm x) &= - g  u(    z),
\end{alignat}
where $u(z)$ and $w(\bm x)$ for the monopole wall are given 
by \eqref{eqn:The harmonic function and the Dirac potential in doubly-periodic space}.
Then the fields for a moving monopole can be obtained by the Lorentz boost.
Keeping the terms of order $q_j^2$, $q_j\bm V_j$ and $\bm V_j^2$, we find
\begin{align}
           \phi ^j(\bm x) &\simeq (g^2 + q_j^2)^{1 / 2}u(z)(1 - \bm V_j^2)^{1 / 2}, \\
           A_\xi^j(\bm x) &\simeq - q_ju(z)V_{j\xi} + g  w(\bm x), \nonumber \\
           A_z  ^j(\bm x) &\simeq - q_ju(z)V_{jz  }              , \nonumber \\
           A_0  ^j(\bm x) &\simeq - q_ju(z)         + g  (wV_j^\xi + \bar wV_j^{\bar\xi}), \nonumber \\
    \tilde A_\xi^j(\bm x) &\simeq - g  u(z)V_{j\xi} - q_jw(\bm x), \nonumber \\
    \tilde A_z  ^j(\bm x) &\simeq - g  u(z)V_{jz  }              , \nonumber \\
    \tilde A_0  ^j(\bm x) &\simeq - g  u(z)         - q_j(wV_j^\xi + \bar wV_j^{\bar\xi}),
    \label{eqn:boosted}
\end{align}
where the scalar potentials are replaced by the Li\'enard-Wiechert potentials with the approximation of the distance
$(r^2 - |\bm r \times \bm V|^2 + O(\bm V^2))^{1 / 2}$ by $r$.

Substituting \eqref{eqn:boosted}  into the Lagrangian for $k = 2$ and 
keeping terms of the second order in $q_1$, $\bm V_1$, $q_2$ and $\bm
V_2$, 
we obtain
\begin{align}
    L_2 &= - m_2 + \frac{1}{2}m_2\bm V_2^2 + q_2(bV_2^\xi + \bar bV_2^{\bar\xi}) \nonumber \\ &
           + \frac{g^2}{2}u(z_2 - z_1)(\bm V_2 - \bm V_1)^2
           - \frac{1  }{2}u(z_2 - z_1)(    q_2 -     q_1)^2 \nonumber \\ &
           + g(q_2 - q_1)\big\{w_{21}(V_2^\xi - V_1^\xi) + \bar w_{21}(V_2^{\bar\xi} - V_1^{\bar\xi})\big\},
\end{align}
where $m_j := v(g + q_j)^{1 / 2}$ is the rest mass of the $j^\mathrm{th}$ monopole wall and $w_{ji} := w(\bm x_j - \bm x_i)$.
Furthermore, expanding $m_j$ and making symmetrization, we obtain 
the total Lagrangian $L_{21}$ as
\begin{align}
    L_{21} &= \frac{vg}{2}(\bm V_2^2 + \bm V_1^2) + \frac{g^2}{2}u(z_2 - z_1)(\bm V_2 - \bm V_1)^2 \nonumber \\ &
            - \frac{v}{2g}(    q_2^2 +     q_1^2) - \frac{1  }{2}u(z_2 - z_1)(    q_2 -     q_1)^2 \nonumber \\ &
            +      b(q_2V_2^     \xi  + q_1V_1^     \xi ) + g     w_{21}(q_2 - q_1)(V_2^     \xi  - V_1^     \xi ) \nonumber \\ &
            + \bar b(q_2V_2^{\bar\xi} + q_1V_1^{\bar\xi}) + g\bar w_{21}(q_2 - q_1)(V_2^{\bar\xi} - V_1^{\bar\xi}).
    \label{eqn:The total Lagrangian of 2-monopole wall}
\end{align}
The Lagrangian may look ill-defined due to the diverging $v$,
however, it can be replaced by 
$v_\mathrm{ren}$ which remains finite (cf.
\eqref{eqn:asymptotic form of the Higgs field}, 
\eqref{8},  
and 
\eqref{eqn:The harmonic 
function and the Dirac potential 
in doubly-periodic space}).
Then the Lagrangian can be divided into the two parts: 
$L_{21} = L_\mathrm{CM} + L_\mathrm{rel}$, where
\begin{widetext}
\begin{align}
    L_\mathrm{CM}  &= \frac{vg}{4}(\bm V_2 + \bm V_1)^2 - \frac{v}{4g}(q_2 + q_1)^2
                    + \frac{     b}{2}(q_2 + q_1)(V_2^     \xi  + V_1^     \xi )
                    + \frac{\bar b}{2}(q_2 + q_1)(V_2^{\bar\xi} + V_1^{\bar\xi}), \\
    L_\mathrm{rel} &= \frac{g^2}{2}\left(\frac{v_\mathrm{ren}}{2g} + \frac{1}{2}|z_2 - z_1|\right)(\bm V_2 - \bm V_1)^2
                    - \frac{1  }{2}\left(\frac{v_\mathrm{ren}}{2g} + \frac{1}{2}|z_2 - z_1|\right)(    q_2 -     q_1)^2 \nonumber \\ &
                    + \left\{\frac{     b}{2} + \frac{\mathrm i\nu g}{8\tau_2}\,\mathrm{sign}(z_2 - z_1)\,(\bar\xi_2 - \bar\xi_1)\right\}(q_2 - q_1)(V_2^     \xi  - V_1^     \xi ) \nonumber \\ &
                    + \left\{\frac{\bar b}{2} - \frac{\mathrm i\nu g}{8\tau_2}\,\mathrm{sign}(z_2 - z_1)\,(    \xi_2 -     \xi_1)\right\}(q_2 - q_1)(V_2^{\bar\xi} - V_1^{\bar\xi}).
\end{align}
\end{widetext}
The center of mass Lagrangian $L_\mathrm{CM}$ would diverge while the
relative Lagrangian $L_\mathrm{rel}$ would converge 
in the limit of $M, N\rightarrow \infty$.
The asymptotic metric of the moduli space 
can be read from the relative Lagrangian.
For convenience, we introduce relative variables by $\xi := \xi_2 - \xi_1$, $z := z_2 - z_1$, $\bm V := \bm V_2 - \bm V_1$ and $q := q_2 - q_1$ and further
replace the electric charge $q$ in $L_\mathrm{rel}$ by the relative phase $\theta$ via the Legendre transformation,
\begin{align}
    L_\mathrm{rel}' = L_\mathrm{rel} + q\dot\theta. \label{eqn:Legendre transf.}
\end{align}
As we will see shortly 
the coefficient of $q\dot\theta$ can be fixed
so that the asymptotic metric has the double periodicity. 
After the Legendre transformation, we obtain the asymptotic metric of the moduli
space in the form of the Gibbons-Hawking ansatz \cite{GiHa},
\begin{align}
    \frac{1}{g}\mathrm ds^2 = U\mathrm d\bm x \cdot \mathrm d\bm x + \frac{1}{U}(\mathrm d\theta + \bm W \cdot \mathrm d\bm x)^2,
    \label{eqn:The asymptotic metric of the moduli space of 2-monopole wall}
\end{align}
where
\begin{align}
    U = \frac{v_\mathrm{ren}}{2} + \frac{g}{2}|z|, \quad
    W_ \xi &= \frac{b}{2} + \frac{\mathrm i\nu g}{8\tau_2}\,\mathrm{sign}(z)\,\bar\xi, \nonumber \\
    W_{\bar\xi} &= \overline W_\xi, \quad
    W_z = 0.
    \label{eqn:Coefficients of the asymptotic metric of the moduli space of 2-monopole wall}
\end{align}
At first sight the metric seems to have a constant shift when we go around the closed cycles on $T^2$,
since $ W_ \xi $ explicitly depends on the coordinate $\bar\xi$.
However we can confirm the double-periodicity of the metric by observing that
the constant shift of  $ W_ \xi $ can be cancelled by the phase shift
due to the necessary $\mathrm U(1)$ gauge transformation
in the twisted boundary conditions
(\ref{eqn:periodic boundary condition 1}) and
(\ref{eqn:periodic boundary condition 2}),
which also determines 
the coefficient of $q\dot\theta$ in (\ref{eqn:Legendre transf.}).
Furthermore, we can also easily check the invariance of the metric
under the modular transformation (\ref{eqn:Modular transformation}). 
Thus our metric   \eqref{eqn:The asymptotic metric of the moduli space of 2-monopole wall} is well-defined
on $T^3 \times \mathbb R$ with local coordinates $(\theta, \xi, z)$. 
Finally the hyperk\"ahler metric \eqref{eqn:The asymptotic metric of the moduli space of 2-monopole wall} allows 
the following local isometries with parameters $(\alpha, \beta, \gamma)$;
\begin{align}
\theta &\to \theta + \alpha +
\frac{\nu g}{4}\,\mathrm{sign}(z)\, 
(\beta  y -\gamma x ), 
\nonumber \\
x &\to x + \beta, \qquad y \to y + \gamma.
\end{align}

It is straightforward to extend the above computation for $k=2$ 
to the case of general $k$.
The total Lagrangian of the $k$ well-separated monopole walls can be
obtained by generalizing (\ref{eqn:The total Lagrangian of 2-monopole
wall}) as follows
\begin{align}
    L_k &= \frac{vg}{2}\sum_{j = 1}^k\bm V_j^2 + \frac{g^2}{2}\sum_{1 \leq i < j \leq k}u(z_j - z_i)(\bm V_j - \bm V_i)^2 \nonumber \\ &
         - \frac{v}{2g}\sum_{j = 1}^k    q_j^2 - \frac{1  }{2}\sum_{1 \leq i < j \leq k}u(z_j - z_i)(    q_j -     q_i)^2 \nonumber \\ &
         +            b\sum_{j = 1}^kq_jV_j^     \xi  +       \sum_{1 \leq i < j \leq k}g     w_{ji}(q_j - q_i)(V_j^     \xi  - V_i^     \xi ) \nonumber \\ &
         +       \bar b\sum_{j = 1}^kq_jV_j^{\bar\xi} +       \sum_{1 \leq i < j \leq k}g\bar w_{ji}(q_j - q_i)(V_j^{\bar\xi} - V_i^{\bar\xi}).
\end{align}
This can be decomposed into the two parts $L_k = L_\mathrm{CM} + L_\mathrm{rel}$, where
\begin{widetext}
\begin{align}
    L_\mathrm{CM}  &= \frac{vg}{2k}\,\Bigg(\sum_{j = 1}^k\bm V_j\Bigg)^2
                    - \frac{v}{2kg}\,\Bigg(\sum_{j = 1}^k    q_j\Bigg)^2
                    + \frac{     b}{k}\,\Bigg(\sum_{j = 1}^kq_j\Bigg)\Bigg(\sum_{j = 1}^kV_j^     \xi \Bigg)
                    + \frac{\bar b}{k}\,\Bigg(\sum_{j = 1}^kq_j\Bigg)\Bigg(\sum_{j = 1}^kV_j^{\bar\xi}\Bigg), \\
    L_\mathrm{rel} &= \frac{g^2}{2}\sum_{1 \leq i < j \leq k}\left(\frac{v_\mathrm{ren}}{kg} + \frac{1}{2}|z_j - z_i|\right)(\bm V_j - \bm V_i)^2
                    - \frac{1  }{2}\sum_{1 \leq i < j \leq k}\left(\frac{v_\mathrm{ren}}{kg} + \frac{1}{2}|z_j - z_i|\right)(    q_j -     q_i)^2 \nonumber \\ &
                    + \sum_{1 \leq i < j \leq k}\left\{\frac{     b}{k} + \frac{\mathrm i\nu g}{8\tau_2}\,\mathrm{sign}(z_j - z_i)\,(\bar\xi_j - \bar\xi_i)\right\}(q_j - q_i)(V_j^     \xi  - V_i^     \xi ) \nonumber \\ &
                    + \sum_{1 \leq i < j \leq k}\left\{\frac{\bar b}{k} - \frac{\mathrm i\nu g}{8\tau_2}\,\mathrm{sign}(z_j - z_i)\,(    \xi_j -     \xi_i)\right\}(q_j - q_i)(V_j^{\bar\xi} - V_i^{\bar\xi}).
    \label{eqn:The relative Lagrangian of multi-monopole wall}
\end{align}
\end{widetext}
On the other hand, the Gibbons-Hawking ansatz for general $k$ can be written as
\begin{align}
    \frac{1}{g}\mathrm ds^2 &= U_{IJ}\mathrm d\bm X_I \cdot \mathrm d\bm X_J
                             + U_{IJ}^{- 1}(\mathrm d\mathit\Theta_I + \bm W_{IK} \cdot \mathrm d\bm X_K) \nonumber \\ &\qquad\qquad\qquad\qquad
                                   \cdot\,(\mathrm d\mathit\Theta_J + \bm W_{JL} \cdot \mathrm d\bm X_L),
\end{align}
where $I, J, K, L = 1, \cdots, k - 1$, and
$\mathit\Xi   _J := \xi   _J - \xi   _k$,
$       Z     _J := z     _J - z     _k$,
$\mathit\Theta_J := \theta_J - \theta_k$ and $\bm X_J := (\mathit\Xi_J, Z_J)$ are relative coordinates
measured by the position of $k^\mathrm{th}$ monopole wall.
By comparing the coefficients of (\ref{eqn:The relative Lagrangian of
multi-monopole wall}) and the sigma model Lagrangian for the  Gibbons-Hawking ansatz, we find:
\begin{align}
     U           _{JJ} &=   (k - 1)\frac{v_\mathrm{ren}}{k} + \frac{g}{2}\sum_{I \neq J}|Z_J - Z_I|,                       \nonumber \\
     U           _{IJ} &= -        \frac{v_\mathrm{ren}}{k} - \frac{g}{2}               |Z_J - Z_I|, ~~\text{($I \neq J$)} \nonumber \\
    (W_     \xi )_{JJ} &=   (k - 1)\frac{b}{k} + \frac{\mathrm i\nu g}{8\tau_2}\sum_{I \neq J}\mathrm{sign}(Z_J - Z_I)\,(\bar{\mathit\Xi}_J - \bar{\mathit\Xi}_I),                       \nonumber \\
    (W_     \xi )_{IJ} &= -        \frac{b}{k} - \frac{\mathrm i\nu g}{8\tau_2}             \,\mathrm{sign}(Z_J - Z_I)\,(\bar{\mathit\Xi}_J - \bar{\mathit\Xi}_I), ~~\text{($I \neq J$)} \nonumber \\
    (W_{\bar\xi})_{IJ} &= (\overline W_\xi)_{IJ}, \quad
    (W_z        )_{IJ}  = 0.
\end{align}

\section{Asymptotic Metric from U(2) Monopole Walls with Singularities}

Finally, we discuss the asymptotic metrics of monopole walls with Dirac-type singularities.
In the case of monopole chains with four-moduli, it is proved that the maximum number of Dirac singularities is four.
Here we derive the inequality for the maximum number of Dirac
singularities of monopole walls by using the spectral curves and the
Newton polygon \cite{ChWa}.
\begin{figure}[htbp]
\centering
\includegraphics[clip,width=4cm,height=4cm,keepaspectratio]{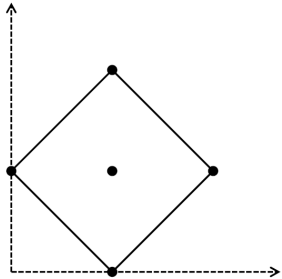}
\caption{
A Newton polygon of an $\mathrm{SU}(2)$ monopole wall with charge $Q_\pm = 1$.
} \label{fig:figure1}
\end{figure}
A spectral curve of a monopole wall is defined by $F_x := \det[V_x(s) -
t]$, where $V_x(s)$ is an integral of $(D_x + \mathrm i\phi)\psi = 0$
in the $x$-direction and $s := \exp[2\pi(z - \mathrm iy)]$.
The spectral curve also induces a spectral polynomial $G_x(s, t) := P(s)F_x(s, t)$, where $P(s)$ is a common denominator of $F_x$.
Then the Newton polygon $N_x$ of $G_x(s, t)$ can be constructed as follows.
Firstly we mark points $(a, b)$ corresponding to the degree of each term $s^at^b$ of $G_x(s, t)$ on an integer lattice.
Then the Newton polygon is a minimal convex polygon including all the marks.
For example, the spectral curves of the $\mathrm{SU}(2)$ monopole walls 
can be written as $F_x(s, t) = t^2 - W_x(s)t + 1$,
where $W_x(s) := \mathrm{Tr}\,V_x(s)$, which leads to the Newton polygon
of an $\mathrm{SU}(2)$ monopole wall with the charge $(Q_-, Q_+) = (1, 1)$ 
as in Figure \ref{fig:figure1}.
In addition, the shape of the Newton polygon is restricted by the boundary data.
For example, the numbers of points on top and bottom edges are equal to
$r_{\pm 0} + 1$, where $r_{\pm 0}$ are the number of 
positive and negative Dirac singularities of a $\mathrm{U}(2)$
monopole wall.
Moreover, there is an important relation between the number of internal points of the Newton polygon, $\mathrm{Int}\,N_x$,
and the dimension of the moduli space $\mathcal M$ 
of the corresponding monopole walls:
\begin{align}
    \dim\mathcal M = 4\,\mathrm{Int}\,N_x. \nonumber
\end{align}

\begin{figure}[htbp]
\centering
\includegraphics[clip,width=7.5cm,height=7.5cm,keepaspectratio]{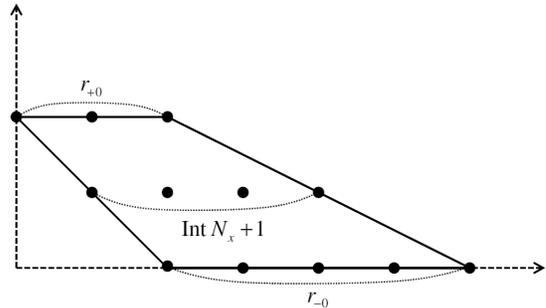}
\caption{
The maximum Newton polygon $N_x$ of a $\mathrm{U}(2)$ monopole wall with $r_{+ 0}$ singularities and $r_{- 0}$ singularities.} \label{fig:figure2}
\end{figure}

Keeping these in mind, the upper limit of the number of singularities of $\mathrm{U}(2)$ monopoles can be easily obtained as follows.
For a given number of internal points, the maximum Newton polygon of monopole walls with singularities must be a trapezoid,
which has height 2 and has length of top and bottom edges $r_{+ 0}$ and $r_{- 0}$ respectively (Figure \ref{fig:figure2}).
From the shape of the Newton polygon, the maximum number of singularities obviously have a relation, $r_{+ 0} + r_{- 0} = 2(\mathrm{Int}\,N_x + 1)$
(which can also be derived by the Pick's formula).
Thus the total number of the singularities $r_0 := r_{+ 0} + r_{- 0}$ is limited as
\begin{align}
    r_0 \leq \frac{1}{2}\dim\mathcal M + 2.
\end{align}
Especially the maximum number of singularities of $k$ well-separated monopole walls is $2k$
because the dimension of the relative moduli space is $4(k - 1)$.
This is consistent with the fact that the maximal number 
of the matter hypermultiplets in the fundamental representation 
is $2k$ in the corresponding $\mathrm{SU}(k)$ super Yang-Mills theory
with eight super charges.

Here we restrict our calculation to the monopole walls with four-moduli,
that is, for $k=2$.
Then the maximal number of the Dirac singularities is $r_0 = 4$.
Since these singularities are stationary and have no electric charge, the metric can be obtained
by simply replacing the vacuum expectation value and the background field by
$v + \sum_{\ell = 1}^{r_0}g_\ell u(    r_{\ell z} -     z)$ and
$b + \sum_{\ell = 1}^{r_0}g_\ell w(\bm r_ \ell    - \bm x)$ respectively,
where $g_\ell$ and $\bm r_\ell := (r_{\ell\xi}, r_{\ell z})$ are 
the magnetic charges and the positions of each singularity \cite{ChKa2002}.
Substituting them into (\ref{eqn:Coefficients of the asymptotic metric of the moduli space of 2-monopole wall}), we obtain
\begin{align}
    U &= \frac{v_\mathrm{ren}'}{2} + \frac{g}{2}|z|
       + \frac{1}{4}\sum_{\ell = 1}^{r_0}g_\ell\left|r_{\ell z} - \frac{z}{2}\right|  \nonumber \\ &\phantom{= \frac{v_\mathrm{ren}'}{2} + \frac{g}{2}|z|}~
       + \frac{1}{4}\sum_{\ell = 1}^{r_0}g_\ell\left|r_{\ell z} + \frac{z}{2}\right|, \nonumber \\
    W_\xi &= \frac{b}{2} + \frac{\mathrm i\nu g}{8\tau_2}\,\mathrm{sign}(z)\,\bar\xi  \nonumber \\ &
           + \frac{\mathrm i\nu}{16\tau_2}\sum_{\ell = 1}^{r_0}g_\ell\,\mathrm{sign}\!\left(r_{\ell z} - \frac{z}{2}\right)\left(\bar r_{\ell\xi} - \frac{\bar\xi}{2}\right)  \nonumber \\ &
           + \frac{\mathrm i\nu}{16\tau_2}\sum_{\ell = 1}^{r_0}g_\ell\,\mathrm{sign}\!\left(r_{\ell z} + \frac{z}{2}\right)\left(\bar r_{\ell\xi} + \frac{\bar\xi}{2}\right), \nonumber \\
    W_{\bar\xi} &= \overline W_\xi, \quad W_z = 0,
\end{align}
where
\begin{align}
    v_\mathrm{ren}' := v - \left(2 + \sum_{\ell = 1}^{r_0}\frac{g_\ell}{g}\right)gC_{M, N}
\end{align}
and we assume $\bm x_1 + \bm x_2 = \bm 0$.

In the correspondence with ${\mathcal{N}}=1$ super
Yang-Mills theory on $\mathbb{R}^3 \times T^2$,
the function $U(z)$ is identified with
the low energy effective coupling,
or the second derivative of the prepotential
on the Coulomb modulus $\mathbb R_{>0}$.

\section{Conclusion}

In this Letter, we have obtained new hyperk\"ahler metrics
whose asymptotic behavior is the ALH type from the moduli
space of monopole walls.  The metric in four dimensions is 
defined on a $T^2 \times S^1$ fibration over $\mathbb R$
and enjoys the modular invariance on $T^2$.
We have also derive the maximal number of the Dirac singularities 
by using  the Newton polygon of the spectral curve. 

One of the next challenges is the low-energy scattering of the monopole walls 
as a geodesic motion on the moduli space.
In the present discussion, the monopoles are 
assumed to be well-separated and hence the collision process is excluded.

In order to obtain a global metric on the moduli space of monopole walls, 
we need some ideas such as the one for the Atiyah-Hitchin metric \cite{AtHi}
for non-periodic BPS $\mathrm{SU}(2), 2$-monopole.
On the super Yang-Mills theory side, the region of well-separated monopoles 
corresponds to the weak coupling region of the moduli space of the Coulomb
branch, where the vacuum expectation values of the scalar fields in the 
vector multiplets are large compared with the dynamical scale of the theory. 
In order to obtain a global metric which is valid on the whole Coulomb branch, 
the inclusion of instanton corrections is crucial. A successful example of 
such computation is the Ooguri-Vafa metric \cite{OoVa}. See also
\cite{GMN} and \cite{Nei} for recent developments.

In the periodic monopoles, 
the monopole scattering has been successfully discussed 
by using the Nahm transform, the spectral curve and the 
corresponding Hitchin equation \cite{HaWa, Maldonado, MaWa}.
These methods could be applied to the doubly-periodic case.


\begin{center}
{\large \bf Acknowledgments}
\end{center}
The authors thank the Yukawa Institute for Theoretical Physics 
at Kyoto University. 
Discussions during the YITP workshop on ``Field Theory and String
Theory'' (YITP-W-13-12) were useful to complete this work. 
MH, HK and DM are supported in part 
by Grant-in-Aid for Young Scientists (\#23740182),
Grant-in-Aids for Scientific Research 
(\#22224001 and \#24540265) and   
Grant-in-Aid for JSPS Fellows, respectively.

\end{document}